\documentclass[acmsmall]{acmart}

\usepackage{booktabs} 
\usepackage{subfig}

\usepackage{multirow}

\setcopyright{rightsretained}
\usepackage{listings}

\begin{document}
\title{Joint Modelling of Cyber Activities and Physical Context to Improve Prediction of Visitor Behaviors}

\renewcommand{\shorttitle}{Joint Modelling of Cyber Activities and Physical Context}



\author{Manpreet Kaur}
\affiliation{
  \institution{RMIT University, Australia\textsuperscript{1}, Tableau Software United States\textsuperscript{2}}
  }
  \email{manpreet882@gmail.com}
\author{Flora D. Salim}
\authornote{Corresponding author}
\affiliation{
  \institution{RMIT University}
  \city{Melbourne}
  \country{Australia}
  }
\email{flora.salim@rmit.edu.au}

\author{Yongli Ren}
\affiliation{
  \institution{RMIT University}
  \city{Melbourne}
  \country{Australia}
  }
\email{yongli.ren@rmit.edu.au}

\author{Jeffrey Chan}
\affiliation{
  \institution{RMIT University}
  \city{Melbourne}
  \country{Australia}
  }
\email{jeffrey.chan@rmit.edu.au}

\author{Martin Tomko}
\affiliation{
  \institution{Melbourne University}
  \city{Melbourne}
  \country{Australia}
  }
\email{tomkom@unimelb.edu.au}

\author{Mark Sanderson}
\affiliation{
  \institution{RMIT University}
  \city{Melbourne}
  \country{Australia}
  }
\email{mark.sanderson@rmit.edu.au}

\renewcommand{\shortauthors}{Manpreet Kaur et al.}

\begin{abstract}
This paper investigates the Cyber-Physical behavior of users in a large indoor shopping mall by leveraging anonymized (opt in) Wi-Fi association and browsing logs recorded by the mall operators. Our analysis shows that many users exhibit a high correlation between their cyber activities and their physical context. To find this correlation, we propose a mechanism to semantically label a physical space with rich categorical information from DBPedia concepts and compute a contextual similarity that represents a user's activities with the mall context. We demonstrate the application of cyber-physical contextual similarity in two situations: user visit intent classification and future location prediction. The experimental results demonstrate that exploitation of contextual similarity significantly improves the accuracy of such applications.
\end{abstract}

%
%




\setcopyright{acmcopyright}
\acmJournal{TOSN}
\acmYear{2020} \acmVolume{1} \acmNumber{1} \acmArticle{1} \acmMonth{1} \acmPrice{15.00}\acmDOI{10.1145/3393692}

\maketitle

\section{Introduction}
Knowledge about consumer behavior is critical for retailers to make personalized recommendations in targeted marketing, improving services, or conduct location prediction. The operators of large indoor shopping malls wish to better understand consumer's behaviors to compete with online retail. Currently, physical retailers primarily gather customer insights by analyzing point-of-sale data. The path a customer took when visiting a mall, how much time they spent at a particular location, or whether they looked for a specific item is information that is typically not available. In contrast, online retailers benefit from rich information about customer activities, including knowledge of Web interaction such as page visits and dwell times. Combined with sales information, such data provides actionable insights that can help retailers improve the online shopping experience of customers. The activities inferred from the data can be exploited to recognize user intent during online shopping. Such an understanding has not been previously explored in physical retail environments.

Malls, museums, galleries, and transport hubs are large heterogeneous environments offering a range of different services: retail, entertainment, information, catering, etc. Increasingly, Wi-Fi networks and Bluetooth$\copyright$ beacons are being introduced into these spaces allowing the logging of movement and information behavior of visitors to such environments. 
Coupled with an understanding of the functions of the different locations in a space (i.e. \emph{physical contexts}) one can ground and classify user behaviors or predict future movements.  
Such an understanding allows the creation and eventual delivery of improved services to visitors.

A person's behavior within a physical space is represented by heterogeneous data, both cyber and physical. In the context of our study, the cyber domain captures a user's interest in the form of queries issued. The physical, associations with Wi-Fi Access Points (APs), captures information related to an area of interest to the user.  We hypothesize that users with \textit{contextual intent} exhibit similarity between their physical contexts and their cyber behavior, i.e. users issue queries related to the context of the physical space. Their cyber-physical behavior reflects what they are interested in. 

To illustrate: consider user $A$ who intends to buy a laptop and compares products online while in the vicinity of a computer store; user $B$, who enters the mall searching for a particular store and follows a trajectory that ends in the store's vicinity; and user $C$, who checks an online footwear size chart while in a store selling shoes. User $A$ is interested in computers, user $B$ is interested in a specific store, and user $C$ is interested in footwear. Such interests can be inferred from the physical context and the combined cyber and physical activity of users.

We present an approach to formulate a correlation between user physical and cyber behavior from heterogeneous data i.e. the Web Query Logs (Cyber) and the Wi-Fi AP association logs (Physical) in order to identify users' interests specific to the physical location. There are number of challenges.
\begin{enumerate}
	\item The \emph{Semantic Labeling} of a physical space. In a mall, this can be done by assigning the category of shops (e.g. Cosmetics, Footwear, Clothing etc) that are in the range of an AP. However, these categories are broad and may not correlate well with a user query.
	\item Therefore, we employ \emph{Semantic Category Expansion} to expand the categories to cover the range of sub categories and products.
	\item To discover the semantic similarity between queries and a physical space we also create a \emph{Contextual Similarity} to map a user query to the representation of categories and relevant sub-categories. 
\end{enumerate}	

The cyber physical contextual similarity shows the users' interests across different semantic categories related to the physical environment and can be used in various applications that involves understanding of user behavior. In our work, we show the use of cyber-physical similarity features in two different applications: Classification of User Visiting Intent and Future Location Prediction. We hypothesize that contextual similarity is a strong indicator of what a user is interested in. It can be helpful in identifying whether a user exhibits high contextual intent with the physical space or is just browsing the area; and, which places the user will visit next.

For behavior classification, the aim is to identify shoppers with high contextual intent, such as users $A$, $B$, and $C$. There are many visitors to a mall for whom their Web behavior and indoor context are \emph{contextually intentless}. Consider user $D$, who visits a mall searching the Web for information about a particular festival occurring in the city, and interleaving these searches with queries about ``lost luggage'' and ``baggage claim''. This user is likely a tourist more focused on the free Wi-Fi than the primary services provided by the mall. While such users clearly have an intent, from the point of view of the mall operator, their visit can be classed as intentless.
We also place in this category `window shoppers', or shoppers with a high-level shopping intent (e.g., `I need to get a present for my brother') that cannot be tied to a particular retailer or category of retailers. While all visitors are potentially of great interest to indoor retailers, we focus our work on detecting contextually intentful customers.

Previous intent recognition relied on examining either physical behavior from Wi-Fi signals, mobile phone sensors, mobile proximity sensors~\cite{Martella2016,you_convenienceprobe:_2014,zeng_analyzing_2015}, or exploiting cyber behavior from online Web browsing and searching logs~\cite{moe2003buying}. To the best of our knowledge, this is the first time a user's contextual intent in an indoor space is inferred from both physical and cyber behavior. 

We also employed user's cyber-physical semantic similarity for future location prediction. Past work~\cite{noulas2012mining} studied the effect of different features on such prediction exploiting Location Based Social Network data. The researchers reported that category of location visited by a user has high impact on prediction accuracy. However, the same is not studied for an indoor setup where movements of a user are captured by Wi-Fi traces. Therefore, we experimented to see if a user's future locations can be predicted accurately by using the semantics of indoor locations visited by the user and query context.

The main contributions of this work are: 
\begin{itemize}
	\item Semantic Categorization used to semantically label a physical space and find the correlation between open text queries and physical semantics;
	\item A Cyber-Physical Contextual Similarity model, used to extract contextual features, including Physical and Cyber activities captured by Wi-Fi AP associations and Web Query logs;
	\item A shopping intent recognition system for user intent recognition, used to classify two broad categories: intentful or intentless. 
	\item An evaluation on the effect of Semantic Context on Future Location Prediction.
\end{itemize}

\section{Background}\label{backgroud}
We categorize our description of past work into five areas.
As mentioned earlier, our main goal is to find cyber-physical semantic similarity from user's cyber behavior captured by Web logs and physical context represented by semantic categories. We then show the application of this similarity in two different applications, User behavior classification and Future Location Prediction.
%




\textbf{Semantic Labeling of Contexts:}
Context is an influential factor in analyzing both human behaviors \cite{pejovic2015mobile} and user intent from mobile information access \cite{church_understanding_2009}. Context is defined as 'any information that can be used to characterize the situation of an entity, where an entity can be a person, place, or physical or computational object.' \cite{abowd1999towards}. 
Semantic labelling of a location context is an important step to identify intent. 
Krumm and Rouhana proposed Placer, which treats semantic labelling as a classification problem based on time, user demographics, and nearby businesses~\cite{krumm_placer:_2013}. 
They found that the demographic information and nearby businesses was helpful in semantic labelling of places, e.g. school, home, and work. 
Later, they proposed an advanced version, called Placer++, which utilizes two more features, the labelled visitors from others and the visit sequences, and found higher accuracy was achieved with these two new features~\cite{krumm_placer_2015}.
Elhamshary et. al. proposed CheckInside, a fine-grained indoor location-based social network,
which utilized check-in data collected from crowd source workers to associate a location with its name and semantic fingerprint.
The researchers claimed CheckInside provides more accurate localization and better coverage~\cite{elhamshary_checkinside:_2014}. 

\textbf{Indoor Behavior Analysis:}
To support real-world, mobile-centric behavioral research, Misra and Balan presented LiveLabs, which is a large-scale mobile testbed for in-situ experimentation~\cite{Misra:2013:LIR:2557968.2557975}.
They also investigated user behaviors when considering whether users are in a group or alone. The researchers found people's mobility patterns, app usage, and propensity to communicate over phones are significantly different across these two scenarios~\cite{jayarajah_need_2015}. 
Martella et. al.~\cite{Martella2016} studied the relationship between indoor visitors and the objects in the case of museum exhibition. Specifically, they deployed energy-efficient mobile proximity sensors to measure the face-to-face proximity between people and objects, and achieved high accuracy of identifying which exhibit a user is facing at short distance~\cite{Martella2016}. 


\textbf{Shopping Behavior Recognition:}
Zeng et. al.~\cite{zeng_analyzing_2015} studied how to determine a shopper's physical behaviors based on channel state information of Wi-Fi signals. They focused on behaviors near shop entrances or within a store, The researchers found the channel information of Wi-Fi signals were a good source to classify these different physical behaviors~\cite{zeng_analyzing_2015}.
Radhakrishnan et. al. presented how to use a smartphone and a smartwatch to segment fine-grained user shopping behaviors: e.g. putting an item in the cart~\cite{Radhakrishnan2016}. Ren et. al. analyzed how people use Wi-Fi to access the Web in indoor retail spaces while navigating through a mall. They found temporal patterns in shoppers' visits and determined that physical context influences user's cyber behaviour~\cite{Ren2015}. Based on these findings, Ren et. al. developed a tripartite location-query-browse graph for contextual recommendations of query, Web content and location, inferred from searching, accessing, and moving behaviors~\cite{Ren2018TKDE}. Using only such behaviours inferred from Wi-Fi logs, Ren et. al. also found strong correlations between behaviours and user demography (e.g. age, gender, income, parental status, visitor types) \cite{Ren2018EPJ}. The work in this paper will build on the contributions from the previous works by Ren et al. \cite{Ren2015, REN:DLog, Ren2018EPJ, Ren2018TKDE} and extends on our previous work on modelling cyber-physical contextual similarity \cite{Kaur2018}.


\textbf{User Intent Recognition:}
Jansen et. al. studied the user intent of Web queries focusing on determining the informational, navigational, and transactional intents~\cite{jansen_determining_2008}. Other work investigated intent based on diary studies by focusing on user mobile information needs~\cite{church_understanding_2009}. The researchers suggested two additional intents: geographical and personal information management. {Chuang-Wen You} et. al. proposed a phone-based system to monitor shopping time in stores classifying user trajectories as either shopping or non-shopping. The researchers utilized spatial and temporal features extracted from both Wi-Fi signals and the accelerometer and digital compass of a phone~\cite{you_using_2011}. Duan and Zhai studied the intent representation problem in the field of entity search, e.g. product retrieval. They proposed a coordinated intent representation by linking the query and entity space collectively. The researcher's focus was on utilizing query terms and product attributes~\cite{duan_mining_2015}. Little research reveals whether shopping intent is detectable in user movement (physical) and query (cyber) behaviors. 

\textbf{Location Prediction:} Exploiting user check-in data from location based social networks to predict/future check-in locations is a well studied topic~\cite{noulas2012mining}. Features used included information on types of places, mobility flows between venues, and spatio-temporal characteristics of user check-in patterns. The work proposed a supervised prediction model, based on linear regression and an M5 model tree.
Another study predicted locations, stay duration, and contact from Wi-Fi and Bluetooth traces~\cite{vu2011jyotish}. For location prediction, the authors use Wi-Fi traces and cluster AP information by exploiting regularity of movement. The study was based on traces collected from fifty participants. 
The former study used the characteristics of venues in terms of categories, the latter did not. 
Recent work that introduced continuous trajectory prediction problem \cite{Sadri2018} also presented a solution based on information-gain to segment multivariate temporal sensor data \cite{SADRI:IGTS}. The study focused on predicting the continuation of user movements, i.e. trajectories, including the sequence of location and departure times for the remainder of the day. Also introduced were two types of trajectories: geographical and labelled.

\textbf{Check-in prediction:} 
Noulas et. al.~\cite{noulas2012mining} analyzed various features from user social network check-in data and found that the types of places users tend to visit (cinema, nightclub, coffee shops etc.) can be highly informative about user mobility preferences. In Location Based Social Networks, venues that user checks in to are labeled with well defined categories, which is not the case with the physical location.
Users' physical movements are captured by 
Wi-Fi AP connections. A location prediction model using Wi-Fi traces was presented in~\cite{vu2011jyotish}. We hypothesize that \emph{Physical Context} (i.e.semantic categories assigned to Wi-Fi APs in a physical location) and \emph{Cyber Context} (i.e user queries context in an indoor space such as shopping centers or museums)  can further enhance prediction results as found in studies on check-ins prediction.

\textbf{Gaps addressed by this research:} 
In our study, we use trajectories, location traces from Wi-Fi APs, labelled with semantic categories of the surroundings. Context information from a user's web query logs can be used to predict future locations. Our evaluation results in successful prediction of less popular locations with higher accuracy then using a model entirely based on mobility flows between locations.

\section{Overview and Dataset Characteristics}\label{cha:system_overview}
\subsection{Research questions}
Given users' \textit{cyber activities} (in terms of web query logs in this instance) and \textit{physical context} (in terms of shop categories), 1) can we enhance the \textit{Physical Context} to determine the correlation between user's \textit{Cyber-Physical context}; 2) how this context can be helpful in applications that involves understanding of user behavior or interests such as for \textit{user intent classification} and \textit{future location prediction}?

Our goal is to determine if there is correlation between user cyber physical behavior and context in a shopping mall. We focus on \emph{user behavior classification} (i.e., shopping intent recognition) and \emph{future location prediction}. We build a model to classify a user trajectory into two broad categories, \emph{intentful} and \emph{intentless}, then we study the effect of semantic or contextual intent on future location prediction.

\subsection{Data Acquisition}\label{sec:data}

We study an anonymized dataset of Internet access, that was captured by an opt-in, free Wi-Fi network in a large inner-city shopping mall in Sydney, Australia. The dataset has a Wi-Fi AP \emph{Association Log} (AL) and a web \emph{Query Log} (QL), collected between September 2012 and October 2013. The APs (around 70) are in the hallway spaces of the mall spread over six levels. The mall contains over 200 stores spanning 29 shop categories, which are defined by the mall operator, Table \ref{table:categories}. 
The locations of the stores and the APs are documented in 2D floorplans. 

The AL contains
1) the association AP ID;
2) the start timestamp of the association;
3) the association duration;
4) the data volume received/sent in this association; and
5) an encrypted persistent user device ID.
The QL contains
1) the query issued by user;
2) the association AP ID at which the query was issued; and
3) the encrypted persistent user device ID. 
The encrypted ID was a hash key based on user registration details and the Wi-Fi MAC address of the device.

\begin{table}
	\centering
	\caption{Mall Operator Defined Categories} \label{table:categories}
	\begin{tabular}{lll}
		\toprule
		Bakeries & Cafe & Cosmetics \\ 
		Costume Jewellery & Delicatessen & Discount Cosmetics \\ 
		Fashion Accessories & Fine Jewellery & General Footwear \\
		Gifts/Souvenirs & Groceries & Gymnasiums \\
		Hair \& Beauty & Home Decor & Men's Fashion \\
		Mobile Phones \& Accessories & Music/Videos/DVDs & Newsagent/Stationery \\ 
		Pad Sites & Repairs \& Maintenance & Restaurant \\
		Small/Major Appliances & Sport & Takeaway \\ 
		Travel & Unisex Fashion & Watches \\
		Women's Fashion & Women's Footwear \\ 
		\bottomrule
	\end{tabular}
\end{table}

\subsection{Query Processing}
\begin{figure}[!bt]
	\centering
	\includegraphics[width=4in]{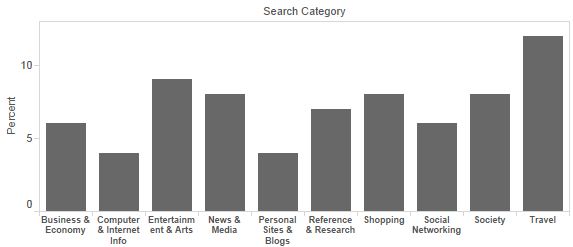}
	\caption{Top 10 Indoor Search Categories}
	\label{fig:searchcategory}
\end{figure}

The queries were grouped into high level categories using the Bright Cloud service (brightcloud.com) to categorize query-click destinations~\cite{ren14:influence}. 
Over 68 categories were found, the distribution of top ten is shown in Figure \ref{fig:searchcategory}. The most popular was Travel, perhaps because the shopping mall is in the center of a popular tourist city. Travelers might be using the free internet. 
We focus on the \emph{shopping} category, around $8\%$ of queries. 


\subsection{AP Association}

\begin{figure}[!bt]
	\centering
	\includegraphics[width=5in]{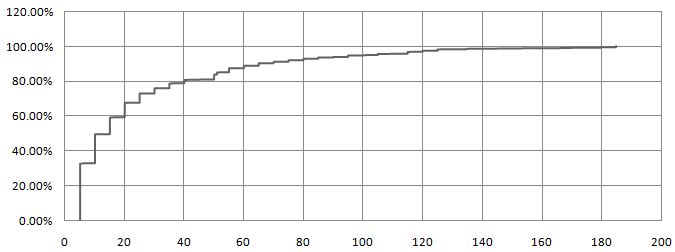}
	\caption{CDF for AP association}
	\label{fig:cdf-aps}
\end{figure}

User movements are captured by AP associations. The associations capture user visits to stores and their passing by a particular location. In order to distinguish between the two we generate a Cumulative Distribution Function (CDF), shown in Figure~\ref{fig:cdf-aps}. 
The AL has a sampling rate of $5$ minutes. The CDF shows around $30\%$ of the associations are found to be $<10$ minutes. Therefore, we only considered a user's association with an AP if the association duration exceeded 10 minutes (two sampling intervals).

\subsection{Web Content-AP Correlation}
\begin{figure*}[!bt]
	\centering
	\includegraphics[width=5in]{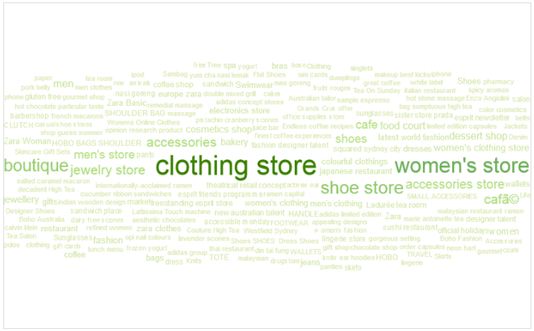}
	\caption{Shopping Center Semantics Word Cloud}
	\label{fig:wordclod}
\end{figure*}

We first extracted semantics labels of the shops (physical context) in the mall from crowdsourced applications including Foursquare, Yelp, and Google places, as shown in Figure~\ref{fig:wordclod}.

\begin{figure*}[!bt]
	\centering
	\includegraphics[width=6in]{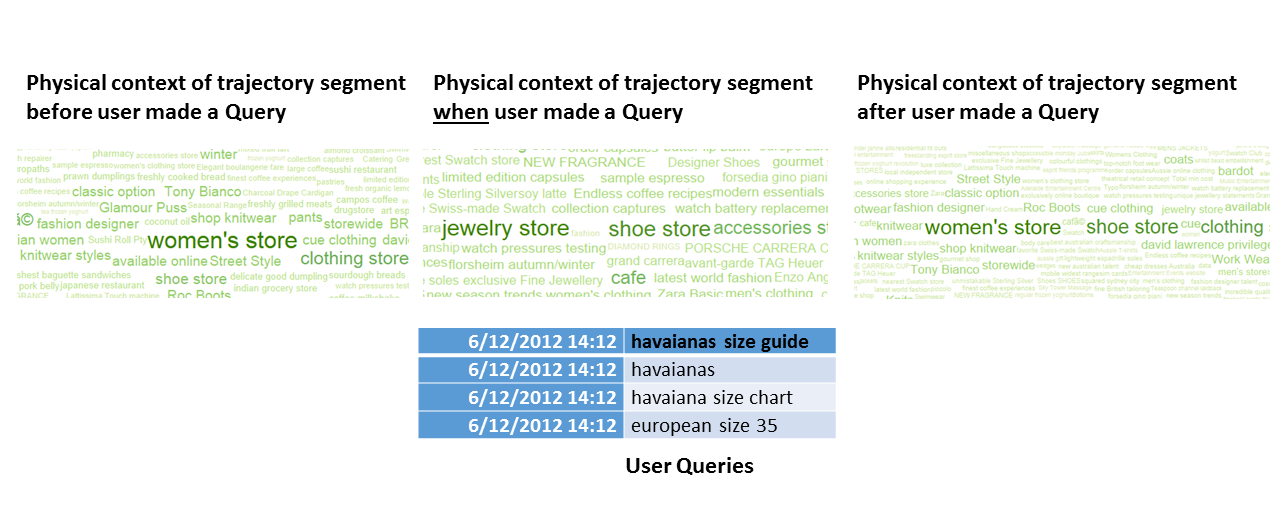}
	\caption{Example of a user search query in co-relation with physical context.}
	\label{fig:usersearch}
\end{figure*}

For each visit of a user, we extracted a trajectory of visited APs. We then analyzed logs by extracting the top user trajectory sessions, as explained in Section \ref{sec:data}. Next, we constructed a sequence of cyber-physical query term sequences that relate the change in information needs with the change in physical context. Figure \ref{fig:usersearch} shows an example of such a trajectory. The intent of the user is first exploring an online footwear size chart when they are close to a shoe store and the category of interest for the user is Footwear. This contextual intent can also be used for future location prediction where the results can be filtered based on the user interest as reflected in queries. 

Thus, we hypothesize that an individual's intent could be constructed by linking their physical behavior (trajectory in terms of shop keywords) and cyber behavior. The challenge is to automatically link the query with the physical context. Intentful query text can contain terms that do not map to currently captured categories. Hence, we propose a \emph{Context Categorization System} (CCS) explained next.

\section{Cyber-Physical Semantic Categorization and Contextual Similarity}\label{cha:Semantic_Categorization}

We define the following:
\begin{definition}
	\emph{Physical Context} is the area of the mall served by the APs, and characterized by the Semantic Categories of Table \ref{table:lat-categories}. 
\end{definition}

\begin{definition}
	\emph{Cyber Context} is a document of entities/categories extracted from users' queries issued in a single visit to the shopping mall.
\end{definition}

In order to identify user intentions, we need to find if their physical trajectories and cyber activities correlate with the physical surroundings. The main challenge is to map open text queries to a small set of category terms that have little or no lexical similarity with each other. For example, a user query may be product name or brand (e.g. \textit{Mascara} and \textit{Ugg Shoes}) which are not in the mall-defined shop category list.

Hence, we propose a system that uses structured information to find intent signals from user queries with respect to physical context, by extending the text of both queries and categories. We gather additional information from DBPedia concepts~\cite{auer2007dbpedia}, extracting categories related to each concept. The extended content representations are then compared. 

We first describe preliminaries, then approach the first task i.e Modelling Physical Space using extended categorical information as an Enrichment of Semantic Categories problem. 
The second task, query extension, we consider it as an Entity Search problem where given a query we try to identify Wikipedia concepts from query text and generate a document of categories related to each entity identified.
At last in Section \ref{sec:Contextual-Sim}, we compare category document generated in Step 2 with each category document in step 1 to generate a vector of similarities to physical context signaling user interests.

\subsection{Preliminaries}\label{sec:preliminaries}
We define some terms. \textit{Documents} are collections of semantic categories.  \textit{Terms} (e.g. shoes, boots) are nouns extracted from queries in the QL. \textit{Entities} are known concepts or resources in DBPedia, which could include specific brand names. \textit{Semantic categorization} is the method to extract \textit{semantic categories} (and the related sub-categories) to represent the physical and query space.
In our work we used two mechanisms to access data from DBPedia: Linked Data and SPARQL. The details of the system are described next.

Here, we describe Semantic Web that we used to do Semantic Category Expansion and finding Cyber Physical Contextual Similarity in section \ref{sec:semantic_web} followed by DBPedia in section \ref{sec:dbpedia}. Then, we show how to access DBPedia in section \ref{sec:accessing_dbpedia}. 

\subsubsection{Semantic Web}\label{sec:semantic_web}
The web has evolved from dumping raw data such as CSV or XML, or HTML tables, sacrificing structure and semantics to linking both documents and the data together so that a person or a machine can explore the web of data. The adoption of linking the data on the web has connected data from diverse domains such as people, companies, books, scientific publications, films, music, television and radio programmes, genes, proteins, drugs and clinical trials, online communities, statistical and scientific data,and reviews enabling users to come up with new applications. The concept of linked data was proposed by~\cite{bizer2009linked} and was achieved by using three descriptive techniques:Resource Description Framework (RDF), Web Ontology Language(OWL) and Extensible Markup Language (XML). \emph{RDF} is a data model that defines the structure and semantics of metadata on the web. It is similar to classical modeling approaches such as entity-relationship or class diagrams, as it structures information about resources in the form of a Triple (subject-predicate-object) expressions. For example: 

\emph{Subject}(Wikipedia page) : \textit{$https://en.wikipedia.org/wiki/Adidas$} of 

\emph{Predicate}(Element) : $http://purl.org/dc/elements/1.1/dct:subject$

\emph{Object}(Category) : \textit{$https://en.wikipedia.org/wiki/Category:Sportswear\_brands$} can be represented in RDF/XML as follows 

\lstset{language=XML,frame = single}
\begin{lstlisting}
<?xml version="1.0"?>
<rdf:RDF xmlns:rdf="http://www.w3.org/1999/02/22-rdf-syntax-ns#"
xmlns:dc="http://purl.org/dc/elements/1.1/"
xmlns:exterms="http://www.example.org/terms/">
<rdf:Description rdf:about="http://dbpedia.org/resource/Adidas">
<dct:subject rdf:resource=
"http://dbpedia.org/resource/Category:Sportswear_brands"/>
</rdf:Description>
</rdf:RDF> 
\end{lstlisting}

The RDF information is structured in xml using RDFS (RDF Schema). In the given example rdf:Description, rdf:about, rdf:resource are parts of RDFS and are standardized following \emph{Web Ontology Language(OWL)} standards.

\emph{OWL} is a language for knowledge representation, a formal way to describe networks and there relationships, where nouns represent objects and the verbs represent relations for example RDFS. This standard representation is followed across all domains and helps to cross link information.

One such example of Semantic Web application is DBpedia that makes the content of Wikipedia available in RDF as explained in next Section.

\subsubsection{DBpedia}\label{sec:dbpedia}
DBpedia, \cite{auer2007dbpedia}, is a project developed on the grounds of Linked Data or Semantic Web that acts as an information extraction framework for Wikipedia. Wikipedia articles is a collection of free text along with  structured information in the form of wiki markup. Such information includes categorisation, images, geo-coordinates, links to external Web pages, disambiguation pages, redirects between pages, and links across different language editions of Wikipedia. DBpedia extracts this structured information from Wikipedia and turns it into a rich information extraction framework representing wiki articles in the form of RDF. The current DBPedia knowledge base as reported by \cite{auer2007dbpedia} describes more than 2.6 million entities, including 198,000 persons, 328,000 places, 101,000 musical works, 34,000 films, and 20,000 companies in the form of 103 million RDF triples that can be used for variety of Semantic Web applications.

\paragraph{Accessing DBPedia Dataset}\label{sec:accessing_dbpedia} DBPedia provides three access mechanisms to its dataset: Linked Data, the SPARQL protocol, and downloadable RDF dumps under GNU Free Documentation License.

\begin{figure*}[!bt]
	\centering
	\subfloat[\label{adidas}]{%
		\includegraphics[width=0.45\textwidth]{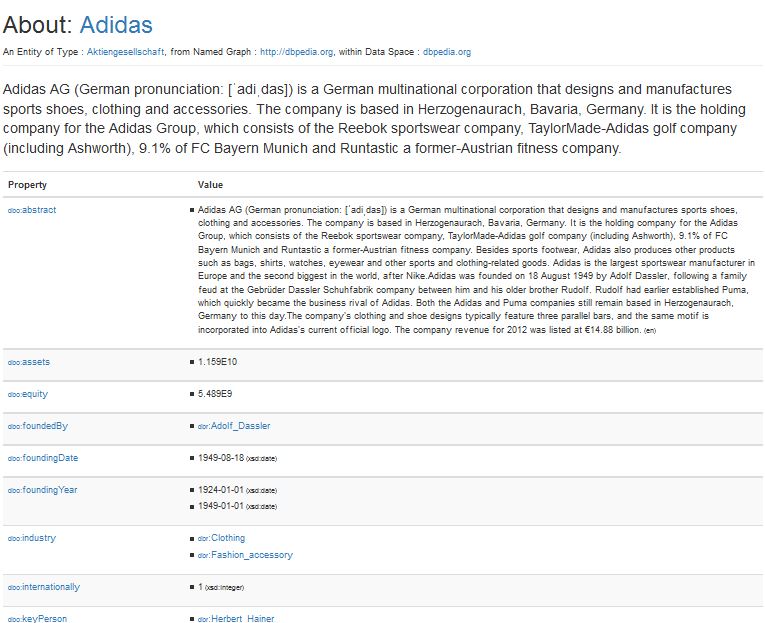}
	}
	\subfloat[\label{adidas2}]{%
		\includegraphics[width=0.45\textwidth]{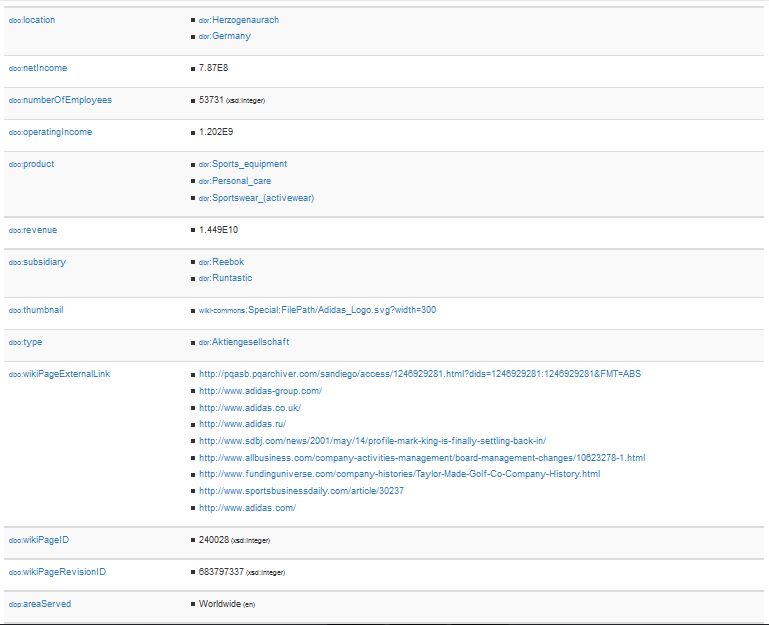}
	}
	\caption{http://dbpedia.org/page/Adidas viewed in a web browser}
\end{figure*}

\emph{Linked Data} is a method of publishing RDF data on the Web that relies on http:// URIs as resource identifiers and the HTTP protocol to retrieve resource descriptions. The URIs return meaningful information about the resource in the form of RDF description. Such a description usually mentions related resources by URI, which in turn can be accessed to yield their descriptions. These URI can be accessed either via web browser as shown in Figures \ref{adidas},\ref{adidas2} or using REST api

\begin{figure*}[!bt]
	\centering
	\subfloat[Sample Query\label{sparql}]{%
		\includegraphics[width=0.6\textwidth]{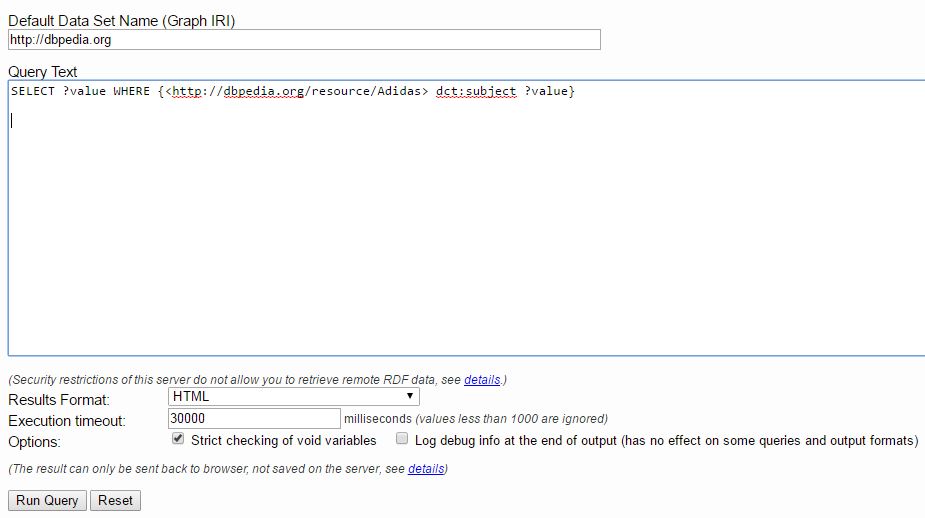}
	}
	\subfloat[Response\label{sparql-response}]{%
		\includegraphics[width=0.3\textwidth]{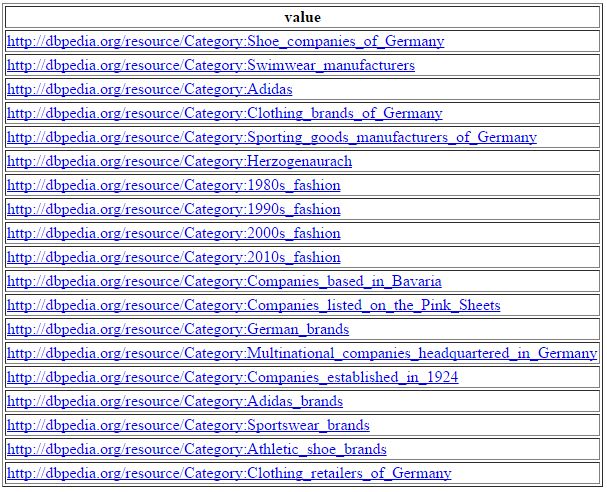}
	}
	\caption{SPARQL DBpedia endpoint}
\end{figure*}

\emph{SPARQL}  is a semantic query language that enables to extract and manipulate data stored in Resource Description Framework (RDF) format. 
Figure \ref{sparql} shows the SPARQL endpoint with the sample query issued to retrieve dct:subject for URI http://dbpedia.org/resource/Adidas and \ref{sparql-response} the response for query issues listing URI's for all the subjects linked to Adidas wiki page.

The last mechanism for accessing data from DBPedia is RDF dumps that are available for download at the DBpedia website and can be used directly.

\subsection{Physical Context}
The semantics of the physical space can be defined as shop categories, as shown in Table \ref{table:categories}. 
However, the categories are too broad for the purposes of user shopping intent recognition. 
Specifically, this study aims to find the correlation between user query with the physical semantics in order to discover the intent of the user. User queries can contain a broad set of terms that can be related to these categories. It is feasible to correlate the user query terms with these categories only using a rich corpus of terms, categories and products that represents shopping center context. 
To generate a corpus that contains a larger range of terms related to shopping context, we use structured information from Wikipedia. Information on Wikipedia is organized by categories and each category has further subcategories forming a tree like structures for the aid of navigation. We exploit this categorical information to enhance semantics and generate a rich corpus of categories. Our hypothesis is, user queries to some extent can be related to Wikipedia categories in order to get an understanding of query intent. Some brand or product-related information is not covered by Wikipedia categories,but most well-known brands and products are covered that are then categorized using relevant Wikipedia categories.
We manually map each semantic category to one of 18 DBPedia categories. We then input each category to our \emph{content categorization system}, which iterates through sub-categories using a depth-first search of up to $\lambda$ levels. We create a document of the iterated categories/sub-categories. Manual tuning led us to set $\lambda=5$ which we found to be an optimal balance between noise and signal. The collection of 18 documents is detailed in Table \ref{table:lat-categories}.

\begin{table}
	\centering
	\caption{Semantic Categories. The number in the brackets denotes the number of sub-category terms, product names, and related terms.}
	\label{table:lat-categories}
	\begin{tabular}{lll}
		\toprule
		Bags (104) & Bakeries (48) & Clothing (183) \\ Coffee (74) & Consumer Electronics (381) & Cosmetics (173) \\
		Decor (188) & Fashion (292)	& Fashion Accessories (203) \\
		Food Retail (91) & Footwear (94) & Home Appliances (174) \\
		Jewellery (153) & Mobile Phones (229) & Restaurants (127) \\
		Retail (214)  & Sports (141) & Watches (123)\\
		\bottomrule
	\end{tabular}
\end{table}

\begin{figure}[bt]
	\centering
	\subfloat[Theoretical Voronoi cells\label{voronoi}]{%
		\includegraphics[width=0.3\textwidth]{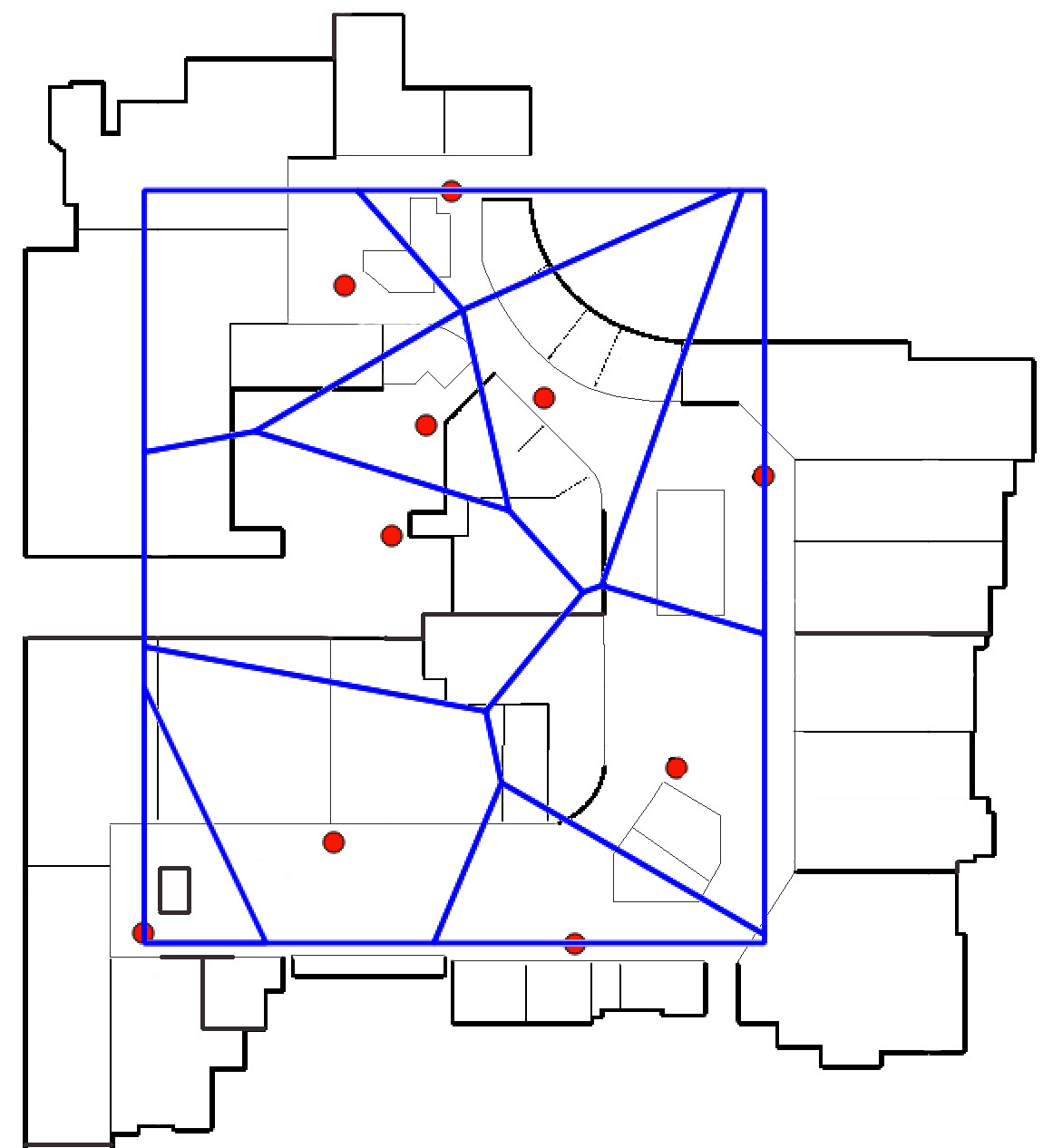}
	}
	\subfloat[Rectified Voronoi cells\label{rectified_voronoi}]{%
		\includegraphics[width=0.3\textwidth]{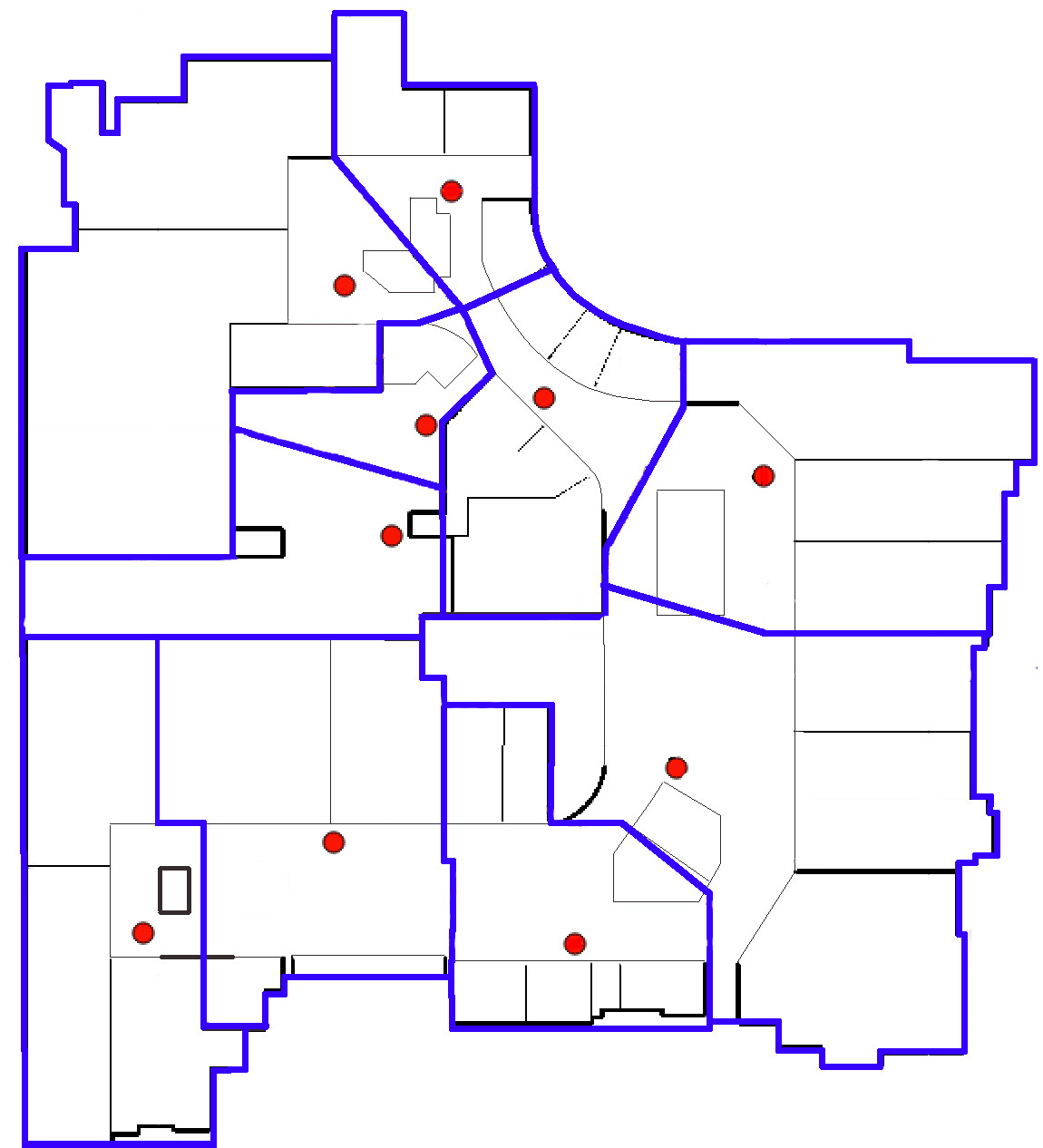}
	}
	\caption{AP coverage regions using Voronoi Cells}
\end{figure}

It is necessary to label each AP with the semantics corresponding to its location in the mall. The physical area of the mall covered by an AP is approximated by a \emph{Voronoi cell}, in which any location is closest to its seed location (the AP) than to any other seed location (other APs)~\cite{bai14:newmethod}, see Figure \ref{voronoi}. 
We manually rectified the cells to match shop frontages and thus better represent physical contexts, see Figure \ref{rectified_voronoi}~\cite{Ren2015}. 
On average, there are 3.67 shops in each rectified cell. The semantic categories of an AP correspond to the categories of each shop in the AP's cell.

\subsection{Cyber Context}

\begin{figure}[tb]
	\centering
	\subfloat[Architecture\label{CCSA}]{%
		\includegraphics[width=0.4\textwidth]{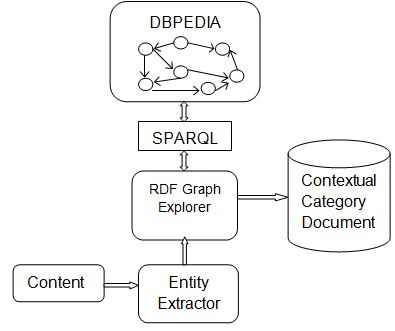}
	}
	\subfloat[Example\label{Example}]{%
		\includegraphics[width=0.4\textwidth]{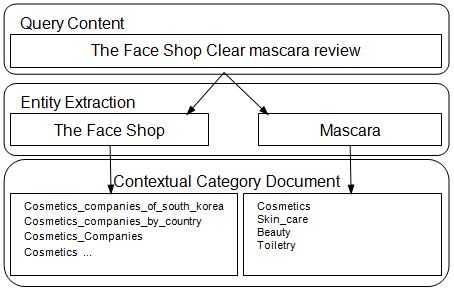}
	}
	\caption{Content Categorization System\label{CCSSystem}}
\end{figure}

Given a query, CCS extracts entities and then gathers contextual DBPedia categories for each entity. The CCS system is shown in Figure~\ref{CCSA}. We describe the components with an example query: ``\emph{The Face Shop clear mascara review}". Note, the process of query categorization is quite similar to \cite{laclavik2015search}

For entity extraction we use Targeted Hypernym Discovery~\cite{Dojchinovski:2013:ECMLPKDD13}, an unsupervised entity discovery and classification system. The system discovered two entities from our example query: \underline{The Face Shop} and \underline{Mascara}.
We then use Graph explorer, which takes a list of entities and looks for resources connected to it via the Simple Knowledge Organization System (SKOS) properties skos:subject and skos:broader. The algorithm iterates through each entity and performs a depth first search on the DBPedia graph. The subject is retrieved only for the main entity discovered and for the broader property. The graph is iterated recursively for $n$ hops to form a contextual category list, which is formed into a document. Figure~\ref{Example} shows the categories that form the document in our example.

\subsection{Cyber-Physical Contextual Similarity}\label{sec:Contextual-Sim}

We now define the contextual similarity between a user's physical movements with what they are looking for online.

We define the physical context as the area of a shop served by a single AP, characterized by latent semantic categories from DBPedia, denoted as $C=\{c_1,c_2,...,c_h\}$, where $h$ is the number of categories. The category documents, from the CCS system, are represented as $D=\{d_1,d_2,...,d_h\}$ composed of subcategories and broader categories for $c_i \epsilon C$. Thus, the physical context for each AP is $P_a:\{ p_{a,1},p_{a,2},...,p_{a,l} \} $, where $p_{a,i} \in C$ for all shops that are located in the Voronoi regions of AP $a_i$. 

We define the physical activity of a user as a trajectory $T=((a_1,t_1),\dots,(a_n,t_n))$, which is a list of tuples of visited AP IDs and the cumulative time of association. We use $a=\{a_1,\dots,a_n\}$ to represent AP, where $n$ is the number of APs user connected to during a single visit to the mall and $t$ to represent time association where $t_k$ is the duration user spent connected to $a_k$ during the visit: $t=\{t_1,\dots,t_k,\dots,t_n\}$. If a user was associated with an AP multiple times in a visit, the total duration of time spent at this AP is stored.

We define cyber context in terms of queries extracted from query logs. During a single session of a user,
we extract all queries of a user represented as $q=\{q1,q2,\dots,qj\}$, where $j$ is the total number of queries extracted from a user. We apply the queries to our CCS system, producing $C_{q_i}=\{c_{q_1},c_{q_2},\dots,c_{q_m}\}$ for each $q_i \in q$. The cyber context is presented as 
\begin{equation}\label{equ-query-con}
Q_c=\bigcup_{i=1}^m c_{q_{i}} = c_{q_{1}} \cup  c_{q_{2}} \cup  \cdots  \cup  c_{q_{m}}.
\end{equation}

The similarity between the physical context $P_c$ and Cyber Context $Q_c$ is calculated in two steps. First, we represent the terms of each document using TF-IDF (Term Frequency - Inverse Document Frequency~\cite{SaltonBuckley1988}) weighting. Then we compute the cosine similarity between physical and cyber using
\begin{equation}\label{con-simi}
cos(d_i,Q_c)=\frac{V(d_i).V(Q_c)}{|V(d_i)||V(Q_c)|}.
\end{equation}
The contextual similarity with Semantic Category $c_i$ represented as $CS(c_i)$ is $cos(d_i,Q_c)$ boosted with Physical Context similarity i.e. time spent at each category denoted as $t_{c_i}$.

\begin{equation}\label{phy-con-simi}
CS(c_i, Q_c)=t_{c_i}*cos(d_i,Q_c)
\end{equation}
where $t_{c_i}>0$ and $cos(d_i,Q_c)>0$.

\subsection{Analysis}
\begin{figure}[h!]
	\centering
	\includegraphics[width=3.5in]{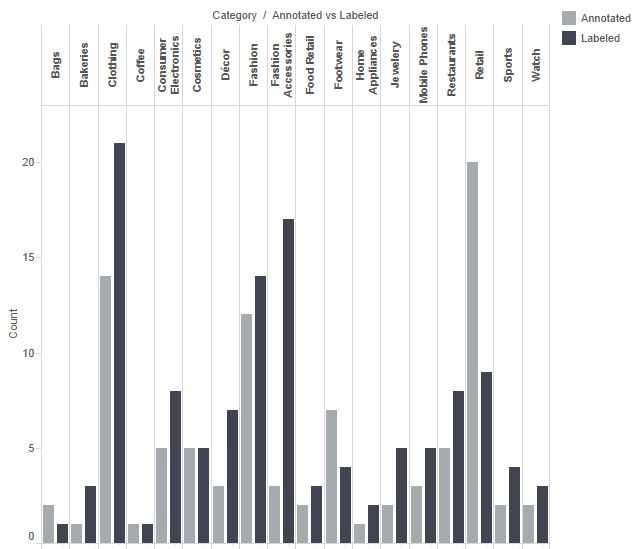}
	\caption{Distribution of manually annotated and labeled categories}
	\label{fig:labeled}
\end{figure}

We examined the cosine similarity between user issued queries and Semantic Categories. We first manually annotated each query with 18 Semantic Categories. The annotation was conducted by three participants who were given a list of queries and a list of semantic categories. They performed the task independent of the contextual similarity model and were asked to label queries with the semantic categories out of the given list. We then compared the annotated categories with the Top-3 categories retrieved by cosine similarity, see Figure \ref{fig:labeled}. To measure the similarity in distribution across the two sets of categories labels, we calculated a Pearson Correlation Coefficient $R = 0.6084$ and found it to be significant, $p$-value $= 0.0073$.

%

\begin{figure}[t]
	\centering
	\includegraphics[width=3in]{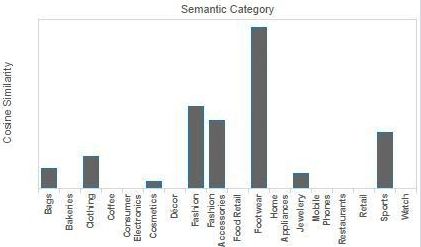}
	\caption{cosine similarity for Cyber Query: Ugg Shoes}
	\label{fig:footwear}
\end{figure}

\begin{figure}[t]
	\centering
	\includegraphics[width=3in]{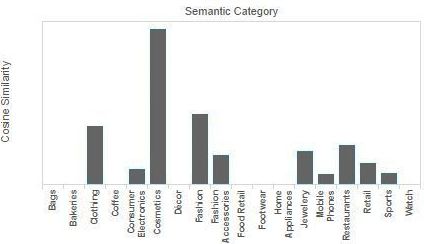}
	\caption{cosine similarity for Cyber Query: the face shop clear mascara reviews}
	\label{fig:cosmetics}

\end{figure}

\begin{table}
	\footnotesize
		\caption{Semantic Categories with Max cosine similarity for sample queries}\label{table:queryCat}
    \centering
	\begin{tabular}{llp{4.5cm}}
		\toprule
		\textbf{Annotated Category} &\textbf{Identified Category}	&\textbf{Query}\\
		\midrule
		Cosmetics	& Cosmetics	&   the face shop clear mascara reviews,\\
		& &  Muk Hair Wax\\
		\midrule
		Clothing	& Clothing	&  Superdry Sale,\\
		& & Emporio Aramani\\
		\midrule
		Fashion	& Fashion	& TopShop Sydney \\
		\midrule
		Footwear & Footwear &  Ugg Shoes\\
		\midrule
		Mobile Phones	&Mobile Phones	& Nokia Lumia 520 reviews \\
		\hline
	\end{tabular}
\end{table}
In Table \ref{table:queryCat} and Figure \ref{fig:footwear}, we see the category with max cosine similarity for the query \emph{Ugg Shoes} (a footwear brand) is the \emph{Footwear}. Figure \ref{fig:cosmetics} shows the cosine similarity for each semantic category for query \emph{the face shop clear mascara reviews}. The max similarity is \emph{Cosmetics}, which is clearly chosen by our annotators. We evaluated query categorization using Accuracy@3 for 217 manually annotated queries. Accuracy was 49.2$\%$, which shows a successful category mapping of the query text.

In our Shopping Intent recognition work (explained  next), we used a similarity distribution across all categories for a given query set per user trajectory. Such a feature vector was shown to improve the accuracy of classification.

\begin{figure}[t]
	\centering
	\includegraphics[width=\columnwidth]{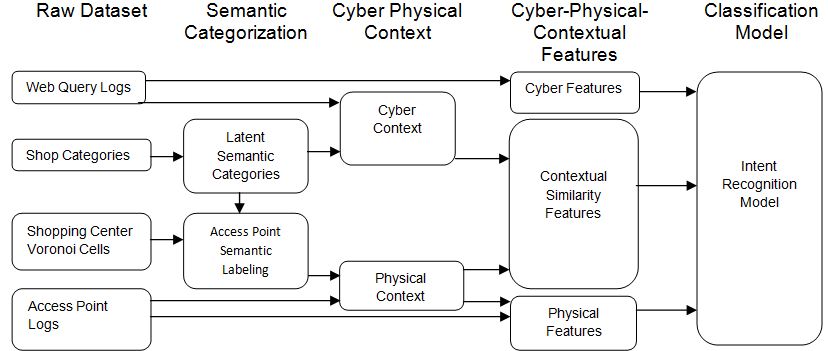}
	\caption{Shopping Intent Recognition System}
	\label{fig:methodology}

\end{figure}

\section{Shopping Intent Recognition System}\label{cha:SIRS}
To review, given an AL, QL, Shop Categories, and Voronoi cells we create an Intent Recognition Model as shown in Figure \ref{fig:methodology}. The first step is to enrich shop categories, provided by the mall operator, with categories from DBPedia. With such enriched categories (stored as documents), we then label each AP with the categories based on the shops within range of the AP's Voronoi cell. After semantic labeling, we determine Physical Activity (trajectories) and Cyber Context for user sessions as recorded in the logs. We then calculate a Cyber-Physical Contextual similarity, which allows us to form Contextual Similarity features that act as input to our Intent Recognition Model along with Physical and Cyber features derived from the AL and QL.
\subsection{Cyber-Physical-Contextual Features}\label{sec:contextual-features}
We investigate an approach for recognizing in-store shopping behavior from an individual's physical movements from Wi-Fi traces and cyber activity from Web queries that users issue.  Our approach rests on the belief that user intent can be identified by correlating their movements with the content they look online. During a typical visit to a shopping center, a shopper uses Wi-Fi either for browsing or when they are looking for some shop or an item they are interested in. If the user who is using Wi-Fi, has a shopping intention, then there is high possibility that they visit some specific category shops and look for related items/category online either to compare the price or for reviews. For example, we show part of the user trajectory in Table \ref{table:1} where user looked for "nest au homeware" online and an association of more that 10 minutes was found with an AP wap032 listed under category "Homeware". We try to correlate this behavior using 3 feature set , Physical, Query and Contextual as given below where we use Trajectory-based Cyber-Physical contextual Similarity  for contextual features. 
Recognition is a binary classifier that labels a user trajectory as Intentful (IF) or Intentless (IL). We examine three feature sets each examined to predict intent.

1) \emph{Physical Activity vs Intent}:
\begin{itemize}
	\item  
F1: Trajectory length: is defined as the number of APs in a user's trajectory;
	\item  
F2: Total duration: how long users spend in the mall in seconds; 
	\item 
F3-F20: Time spent per shop category: the distribution of the total duration over shop categories.
\end{itemize}

2) \emph{Cyber-Physical activity vs Intent}:
\begin{itemize}
	\item 
F1-F20: As defined above
	\item 
F21: Number of queries.
\end{itemize}

3) \emph{Contextual Features vs Intent}:
\begin{itemize}
	\item 
F22-F39: $CS(c_1)$-$CS(c_{18})$ - Contextual Similarity of User's Cyber-Physical activity with Semantic Category documents i.e. $d_1$-$d_{18}$;
	\item F40:  
Max Contextual Similarity - is $max(CS(c_1):CS(c_{18}))$;
	\item 
F41: Sum of $CS(c_1):CS(c_{18})$;
	\item 
F42: the cosine similarity of categories extracted from user issued queries in a single visit with the list of over stores in the mall;
	\item 
F43: the cosine similarity of categories extracted from user issued queries in a single visit against a list of keywords/categories extracted from crowdsourced Web applications including Foursquare, Yelp and Google places for stores in the mall. 
\end{itemize}

\begin{table}
\footnotesize
	\caption{User Trajectory}
	\label{table:1}
	\begin{center}
		\begin{tabular}{lll} 
			\toprule
			\textbf{Wi-fi AP} & \textbf{AP Semantics} & \textbf{Query}  \\ 
			\midrule
			\multirow{3}{4em}{wap030} &  Restaurant &  nest au homeware  \\ 
			& Cafe & \\ 
			& Groceries &  \\ 
			\midrule
			\multirow{3}{4em}{wap032} &  Homeware &  \\ 
			& Clothing & \\ 
			& Footwear &  \\ 
			\midrule
			\multirow{3}{4em}{wap009} &  Clothing &    \\ 
			& Footwear & \\ 
			\midrule
			$\cdots$ & $\cdots$	& $\cdots$	\\ [1ex] 
			\bottomrule
		\end{tabular}
	\end{center}
\end{table}

\subsection{Intent Recognition Model}\label{sec:IRM}

As most of our cyber-physical-contextual features are independent of each other, we deploy a Decision Table/Na\"ive Bayes (DTNB) hybrid classification method \cite{hall2008combining} to perform the Intentful and Intentless classification. The method selects the deterministic features for recognizing intent from a range of input features. We examine how each method performs.

\paragraph{Decision Table (DT) Model}

Given a set of labeled instances as a training sample, an induction algorithm creates a decision table with default rule mapping to the majority class. The DT model has two main components:
\begin{itemize}
	\item Schema: a set of features selected by maximizing cross-validated performance using forward search.
	\item Body: a multiset of labeled instances.
\end{itemize}
Each instance consists of a value for each of the features in the schema and a value for the class.

For label assignment to an unlabeled instance $I$ by a DT model classifier, let $L$ be the set of labeled instances in the model
matching a given instance $I$.  There is a match between $2$ instances if the features in the schema are same. 
If $L=0$, the DT model returns the majority class, otherwise it returns the majority class in $L$. 

\paragraph{Na\"ive Bayes (NB)}
This widely used classifier takes the following form:
\begin{equation}\label{bayes-theorem}
p(l_i|f_i)=\frac{p(f_i|l_i) p(l_i)}{p(f_i)},
\end{equation}
where $l_i$ is a class label and $f_i$ is a contextual feature; $p(l_i,f_i)$ is the probability of $f$ in $l_i$; $p(f_i|l_i)$ is the probability of $f_i$ given class $l_i$; $p(l_i)$ is the probability of occurrence of class $l_i$ and $p(f_i)$ is the probability of occurrence of feature $f_i$.

Considering the features are defined from physical and cyber perspectives, we assume that they have an independent distribution. Thereby Eq.~\ref{bayes-theorem} becomes:
\begin{equation}\label{bayes-theorem-if}
p(f|l_i)=p(f_1|l_i)*p(f_2|l_i)*...*p(f_n|l_i).
\end{equation}

In a classification task, given a feature set $f=\{f_1,f_2,...,f_n\}$ for binary classification of $\{l_i,l_j\}$, NB labels an instance as class $l_i$ if its posterior probability is higher than the other class, namely $p(f|l_i)>p(f|l_j)$.

\paragraph{A DTNB Bayes Hybrid model}
The model is a simple Bayesian network in which
the DT represents a conditional probability
table~\cite{hall2008combining}. The algorithm for learning the combined model (DTNB) works in a similar way as that of stand-alone DT. It partitions the feature set into two disjoint subsets: one for the DT, the other for NB. Then, it uses forward selection, where, at each step, selected attributes are modeled by NB and the remainder by the DT. 

The class probability of the DT and NB are then
combined to generate overall class probability estimates.
Assuming $f_{DT}$ is the set of features in the DT and $f_{NB}$ the
one in NB, the overall class probability is computed as
\begin{equation}\label{eq:dtnb}
P(l_i|f) = a * P_{DT} (l_i|f_{DT}) * P_{NB}(l_i|f_{NB}/P(l_i))
\end{equation}
where $P_{DT}(l_i|f_{DT})$ and $P_{NB}(l_i|f_{NB}) $ are the class probability
estimates from the DT and NB respectively, $a$
is a normalization constant, and $P(l_i)$ is the prior probability of the class label $l_i$. 

\subsection{Future Location Prediction}\label{cha:location_prediction}

We now investigate the following question: 
Given a user's physical and cyber activities, is the semantic content of queries of value for location prediction? 
From initial analysis of the data (Wi-Fi APs association and Web logs), we found that user queries in a shopping center are somewhat indicative of their interests for that particular visit. For example, user $A$ enters the mall and searches for a particular store and follows a trajectory that ends in the vicinity of that store. Here, we try to find if the contextual information can be exploited for future location prediction in an indoor space by using Collaborative Filtering as the baseline prediction model.
We first formulate the problem, then describe the methodology, and detail experiments.

\subsubsection{Problem Formulation}\label{sec:problem_form}
Given a list of $m$ user trajectories $T = \{t_1, t_2, . . . , t_m\}$ and a list of $n$ APs $A = \{a_1, a_2, . . . , a_n\}$. Each user trajectory $t_i$ has
a list of APs $A_{t_i} \subset A$, which the user has connected to in the order of association time. The user issues a set of $n$ queries $Q =\{q_1,q_2,...,q_n\}$. We can, therefore, calculate the likelihood of visiting an un-visited AP $a_j \notin A_{t_i}$ for the trajectory $t_j \in T$.

\subsubsection{Methodology}\label{sec:method}
We used Item-based collaborative filtering as the baseline model.
For the recommendation algorithms, we deploy both 
User-Based Collaborative Filtering and Item-based Collaborative Filtering~\cite{DBLP:journals/concurrency/RenLZ15}
with Contextual Similarity defined as follows.  

\paragraph{User-Based Collaborative Filtering}
This method solves the recommendation problem from the users' perspective. It firstly identifies the neighbors of a target user (i.e., they either rate different places similarly or they tend to visit similar places), then tries to aggregate the neighbours' opinion to estimate what the target user may like. 
This technique is widely used in practice for item recommendation and location prediction.

\paragraph{Item-Based Collaborative Filtering}\label{sec:i-i}
The item-based approach investigates the set of locations a target user has rated or visited and computes their similarity to the target (un-visited) location. It then selects the $k$ most similar locations ${p_1, p_2, . . . , p_k }$ for recommendation. 

\paragraph{Similarity Computation}
The similarity between users or items plays a key role in both algorithms. Given the rating vectors of item $i$ and $j$, we aim to find how similar these two items are rated by a set of users and then to calculate the similarity $s_{i,j}$ between them. There are a number of different metrics to these similarities. But as we have a binary vector -- where 0 represents a location not visited and 1 represent a location visited by the user -- we chose \emph{Jaccard Similarity}. Given two items $i,j$ represented as binary vectors $I,J$ in $m$-dimensional user space. The similarity ${s_{i,j}}$ is computed as follows:

\begin{equation}
{s_{i,j}} = J(I,J) = \frac{|I \cap J|}{|I \cup J|} = \frac{|I \cap J|}{|I| + |J| - |I \cap J|}.
\end{equation}

Given a user trajectory $t_i$, a similarity value $s_{a_i,a_j}$ is calculated for all items $a_i \in t_i$ (locations visited by the user) and $a_j \in |A-t_i|$ (all other locations not visited by the user). 
Thus, we obtain the similarities between all visited and all un-visited locations. 
For each un-visited location $a_j$, we take the average of its similarities to each $a_i \in t_i$ as its estimated similarity to the target user: $s(i,j)=1/n \sum_{i=1}^{i<n} s_{i,j}$.

For top $k$ prediction, we sort the similarity vector $s$ and retrieve top-k locations. This provides us with a set of items that are most likely to be visited by the user based on the historical locations visited. Next, we weigh this prediction score for each AP using a contextual similarity of queries issued by the user.

%
%

The similarity between the physical and cyber Context $Q_c$ is calculated in two steps. Firstly, the cosine similarity between each document $d_i \in D$ and $Q_c$ (TF-IDF vector) is measured.
This generates a similarity vector $CS$ of size $h$ where $CS_i$ is the similarity of query document $Q_c$ with category document $d_i$. Secondly, we generate a dot product of the physical context vector $P_{a_i}$ for $a_i \in A$ with the similarity vector $CS$ that represents the user query context corresponding to each $a_i$ as follows:
\begin{equation}
SS = P_{a_i}\cdot CS=\sum_{j=1}^{h}P_{a_i,j}\cdot CS_j = P_{a_i,1}\cdot CS_1 + P_{a_i,2}\cdot CS_2 + ..... + P_{a_i,h}\cdot CS_h
\end{equation}
where $h$ is the number of categories.

The semantic similarity vector $SS$ is used to weigh the item-item similarity score by taking the product of $JS_i$ and $SS_i$, 
where $i$ denotes the index of an AP $a_i$:
\begin{equation}
weightedSimilarity(JS_i,SS_i) = JS_i*SS_i.
\end{equation}

The prediction is given by sorting the weighted similarity score and extracting top $k$ items.

\section{Experiments}\label{experiments}

We focus our experiments on a subset of \emph{complete} user trajectories with associated user queries. A complete trajectory is one where the start and end points correspond to entry/exit points of the mall. Such trajectories must connect at least three APs. Out of $6784$ total trajectories, we identified 176 that are complete.
Four annotators, without in-depth knowledge of the experiment, manually categorized the trajectories into \emph{intentful} (48) and \emph{intentless} (128), with 100\% inter-annotator agreement. The annotators inspected the queries and marked them as \emph{relevant} if the content was deemed related to the environment of the shopping centre. A session was labelled as \emph{intentful} if at least one of the queries was \emph{relevant}, and \emph{intentless} otherwise.

To evaluate the classification models we used: 
$Accuracy\%$, the percent of correct classifications;
$Precision$, the percent of correct positive classifications;
$Recall$, the percent of positive instances correctly classified; and
$F-Score$, a weighted harmonic mean of Precision and Recall.

To evaluate location prediction, we used: 
\emph{Accuracy@k}, the number of correct locations predicted over $k$, which is the total no. of locations predicted; and 
\emph{MRR}, the ranking of first correct location, 
$MRR=\frac{1}{n}\sum_{i=1}^{n}\frac{1}{R_i}$,
where $n$ is the no. of prediction results and $R_i$ is the rank of first correct predicted location for trajectory $i$.


%
%
%
%
%
%

\begin{table}[h]
\footnotesize
	\begin{tabular}{l|l|l|l|l|l}
		\toprule
		\textbf{Features}	& \textbf{Method}	& \textbf{Accuracy \%} & \textbf{F-Score} & \textbf{Precision} & \textbf{Recall}\\
		\midrule
		& NB   & 63.06&	0.59&	0.56&	0.63\\
		Phy & DT & 72.73&	0.61&	0.53&	0.73\\
		& DTNB   &72.73	&0.61&	0.53&	0.73\\
		\midrule
		& NB  & 63.06&	0.59&	0.56&	0.63\\
		Phy + Cyb&DT &78.41	&0.73&	0.81&	0.78\\
		& DTNB   &78.41	&0.73	&0.81&	0.78\\	
		\midrule
		& NB  & 73.29&0.68&	0.69&0.73\\
		Cont &DT &76.13	&0.73&	0.74&	0.76\\
		& DTNB   &76.7&0.75	&0.75&	0.77\\	
		\midrule
		& NB   &69.32&	0.66&	0.65&	0.69\\
		Phy + Cont& DT &76.14&	0.73&	0.74&	0.76\\
		& DTNB   &76.14	& 0.74&	0.74&	0.76\\
		\midrule
		& NB   &69.32&	0.66&	0.65	&0.69\\
		Phy + Cyb + Cont& DT &78.41	& 0.75&	0.79&	0.78\\
		& DTNB   &\textbf{81.25}&	\textbf{0.8}&	\textbf{0.8}&	\textbf{0.81}\\
		\hline
	\end{tabular}
	\centering
	\caption{Intent Recognition Results}
	\label{table:results}

\end{table}

\subsection{Results on Intent Recognition}
As shown in Table \ref{table:results}, the DTNB hybrid classifier always performs comparably or better than DT and NB. The best accuracy of 81.25\% is achieved with DTNB on all Cyber-Physical-Contextual features. 
\begin{table}[h!]
\footnotesize
	\begin{tabular}{c|c|c|c}
		\toprule
		\textbf{Features}	& $df$ & $t$ & $p$-value\\
		\midrule
		Phy - Phy+Con  & 11	&4.7833&	0.0006\\	
		\midrule
	Phy+Cyb - Phy+Cyb+Con & 11 &	3.1747	&0.0008\\	
		\bottomrule
	\end{tabular}
	\centering
	\caption{Paired t test}
	\label{table:ttest}
\end{table}
The results show that the increase in performance with contextual features is statistically
significant ($p$-value = $0.0006$ for Physical vs Physical + Contextual, and $0.0008$ for Cyber vs Cyber +Cyber + Con. 
The paired $t$-test statistics are shown in Table \ref{table:ttest}. 

\subsection{Results on Future Location Prediction}
\label{sec:prediction_experiments} 

\paragraph{Dataset}\label{sec:exp}
We performed a prediction experiment on 994 complete and partial trajectories where at least one query was issued. 
We then partitioned 325 trajectories into training and test trajectories. The partition point is the AP where the user issued their first query and the rest of the trajectory are used for evaluating prediction results to see if the semantic context of queries with respect to physical locations helps in improving prediction results.
We then used 669 full trajectories and 325 partitioned train trajectories to generate a collaborative filtering matrix in order to get Top-10 prediction results for the 325 partitioned test trajectories using simple Item-Item similarity method (denoted as \textit{i-i})
 and the improved similarity score computation using Weighted Similarity (denoted as \textit{i-i-w}). 



\begin{figure}[!bt]
	\centering
	\includegraphics[width=0.75\textwidth]{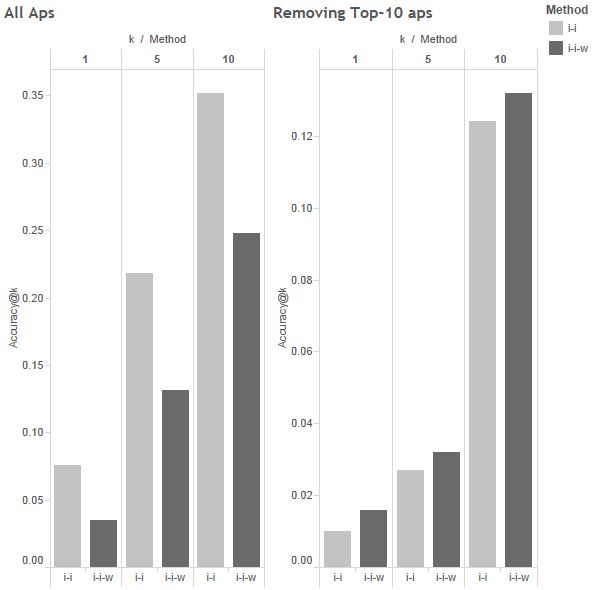}
	\caption{Prediction Results}
	\label{fig:prediction}
\end{figure}

\begin{figure}[!bt]
	\centering
	\includegraphics[width=0.75\textwidth]{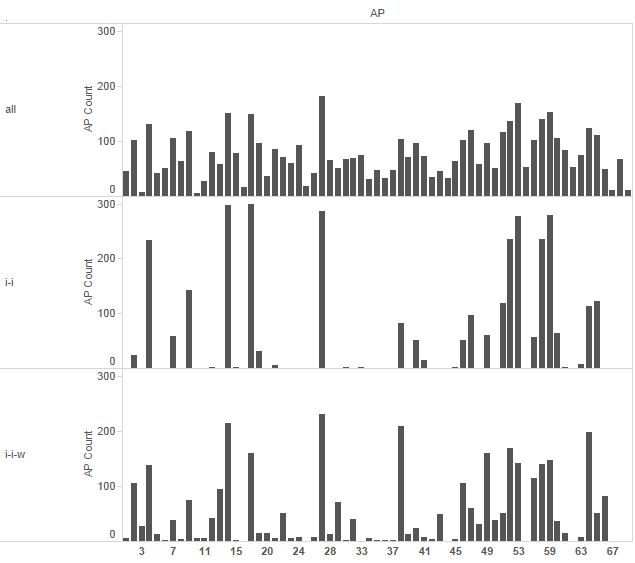}
	\caption{Prediction Results}
	\label{fig:prdiction}
\end{figure}

\begin{figure}[!ht]
	\centering
	\includegraphics[width=3in]{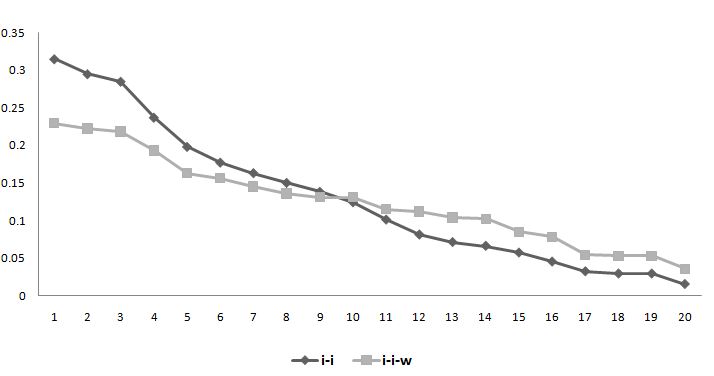}
	\caption{Sensitivity Analysis of Accuracy@10 by removed top-n APs, $n$ ranging from 1 to 20}
	\label{fig:sensitivity}
\end{figure}

\begin{figure}[!ht]
	\centering
	\includegraphics[width=2in]{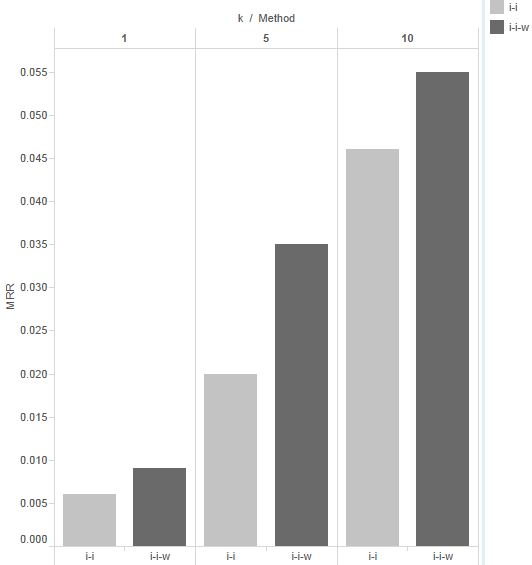}
	\caption{Mean Reciprocal rank for k predictions, $k$ = 1,5,10}
	\label{fig:mrr}
\end{figure}
The bar chart on the left in Figure~\ref{fig:prediction} shows results with no improvement in accuracy using contextual similarity weight. We then generated a chart of predicted APs on the $x$-axis and a count of APs on the y-axis in the test set (all) along with predicted APs using \emph{i-i} and \emph{i-i-w} methods as shown in Figure~\ref{fig:prdiction}. We see that the \emph{i-i-w} method (bottom graph in Figure~\ref{fig:prdiction}) performs well at predicting less popular APs. This can be because less popular locations might be semantically similar.

To assess the correctness of our assumptions based on the chart, we performed a sensitivity analysis on Accuracy@10 by removing the top 20 APs. Figure~\ref{fig:sensitivity} shows that \emph{i-i-w} consistently outperforms \emph{i-i} after removing some of the popular APs. We then measured Accuracy@k for $k=\{1,5,10\}$ after removing the Top-10 APs from the test set (Figure~\ref{fig:prediction}). We see an improvement in accuracy for item-item weighted (i-i-w) compared to item-item (i-i) where accuracy increases with an increase in the number of predictions ($k$). The improvement in accuracy is statistically significant between \emph{i-i} and \emph{i-i-w} ($p$ = $0.0188$,  two-tailed paired $t$-test \cite{hsu2008paired}). 
We also used \emph{MRR} to evaluate the ranking of first correct location predicted using \emph{i-i} and \emph{i-i-w} for top-k prediction, $k={1,5,10}$. As shown in Figure \ref{fig:mrr}, MRR for \emph{i-i-w} is better then \emph{i-i}. We thus conclude that contextual similarity improves prediction of less popular locations with better ranking as well. 

%
\section{Discussion}\label{cha:discussion}
For behavior recognition, we first performed an experiment to classify users' shopping intent as \emph{intentful} and \emph{intentless}, by using users' Cyber-Physical-Contextual activities captured by a Wi-Fi APs association log and a Web query log. We proposed a Shopping Intent Recognition System, which includes Semantic Categorization to semantically label a physical space and find the correlation between open text queries with the physical semantics. We also described Cyber-Physical Contextual Similarity model to extract contextual features including Physical and Cyber activities captured by Wi-Fi AP and Web Query logs. Finally, we detailed a User intent recognition system to classify a user's intent.

We showed that the proposed contextual features significantly improved the accuracy of intent recognition models where we used Decision Tables, Na\"ive Bayes and a Decision Table Na\"ive Bayes hybrid model.
The models were applied to a set of 176 real user trajectory sessions in an indoor shopping mall. The DTNB achieved the best performance, compared to DT and NB over all feature sets. The maximum classification accuracy of 81.25\% was achieved by using DTNB on all feature sets (Cyber-Physical-Contextual). We also note that in the entire query log dataset for the target environment, only 8\% of the queries belong to the broad shopping semantic categories from Table~\ref{table:categories}. If this set of queries was larger, the similarity detected may not be conclusive. But because this category of queries represents a small subset of the overall query activity of indoor mall users, the high level of similarity detected between the semantics of the physical context and those of the query activity are strong indicators of intentful activity.

We further performed an experiment to study the effect of Cyber-Physical Contextual Similarity on location prediction using Collaborative Filtering. Using contextual features as a weight to the probability likelihood calculated from collaborative filtering model improves the accuracy of prediction of less popular locations.%

\begin{figure*}[!bt]
	\centering
	\includegraphics[width=5in]{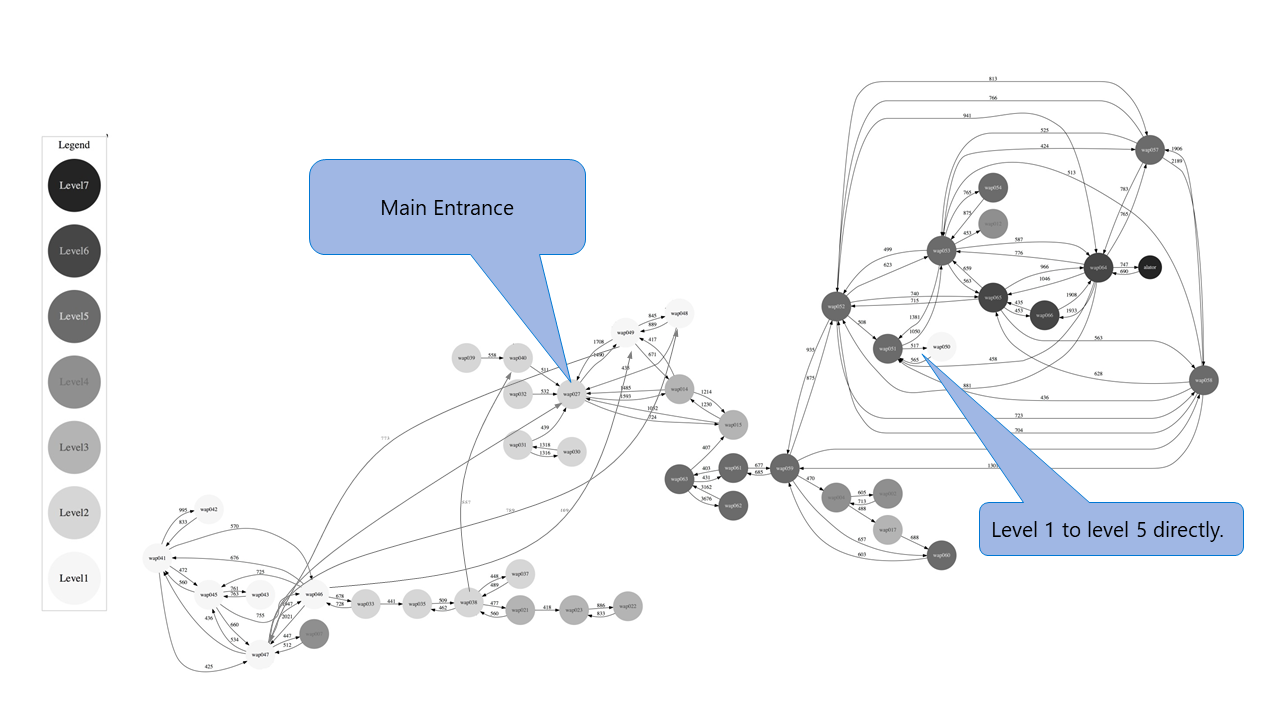}
	\caption{Movement Flows between Wi-Fi Access Points across Floors}
	\label{fig:APFlows}
\end{figure*}

\textbf{Limitations.}
This paper has largely focused on the semantic expansion of spatial features, but more can be done on embedding temporal features. The data evidently reveals that some trajectory behaviours are specific to certain visiting intent (e.g. going for a lunch) on different temporal contexts, such as shown on Figure \ref{fig:APFlows}, where there are direct movements between the first floor to fifth floor during noontime and lunch break period. The fifth floor is where the food court is located. If we also extract the temporal features and expand the semantic representation, this could potentially boost the predictability of the intent.

Further, these features could also be used further for profiling users \cite{Ren2018EPJ}, useful for providing a personalised model to predict the visiting intent and next location indoors. 

In addition, sequential or continuous behaviours of the whole trajectory in a session is not yet incorporated in the CPS model of this paper. From our earlier paper, we have observed repeating and habitual behaviours of returning visitors are observed across the history \cite{Ren2015}, which could be used for continuous trajetory prediction \cite{Sadri2018} for example, or intelligent notification, shopping assistant, or recommendation \cite{Ren2018TKDE} purposes. 

Finally, a complete semantic expansion and embedding of cyber, physical, social behaviours can be done in the future. Recent work on graph embedding, especially node and relation embedding \cite{wang2020relation} can be used to deal with the data sparsity and cold start problems in this data (or similar datasets) when high-dimensional features are combined across multiple sources or domains. Such an embedding could also generate more effective recommendation results.

\section{Conclusion and Future Work}\label{cha:conclusion}
We proposed a semantic enrichment and contextual similarity model that deals with the major challenge of mapping semantic similarity across two different domains: cyber and physical behaviours. 
We show that using this combined contextual similarity further improves the accuracy of both intent recognition and location prediction with respect to the use of cyber/physical features in isolation.

There are some limitations to our work that can be improved in the future. Firstly, we performed the Shopping Intent Recognition on a small dataset, as manual labeling of trajectories and the respective query sets was required. Future experiments with crowdsourced labelling of much larger datasets are envisaged. Secondly, we only studied the effect of contextual features on location prediction. Future works should investigate the effect of other features, such as time spent at an AP captured through other distance metrics (e.g., cosine similarity or Pearson correlation). Finally, as we characterize users based on their detailed cyber activities and physical contexts, the construction of group profiles based on demography and visiting patterns will be further investigated. 

\section{Acknowledgements}
We acknowledge the support of Australian Research Council (ARC) Linkage Project \textit{LP120200413}. 
Flora Salim would also like to acknowledge the support of Alexander von Humboldt Foundation.
We also want to thank the referees of the previous versions of
this paper for their useful suggestions.

\bibliographystyle{ACM-Reference-Format}
\bibliography{Bib/strings,Bib/main}

\end{document}